\documentclass{aa}  
\usepackage{graphicx}
\usepackage{natbib}
\bibpunct{(}{)}{;}{a}{}{,}
\usepackage{txfonts}
\begin{document}
\title{Evidence for metallicity spreads in three massive M31 globular clusters\thanks{Individual photometric measurements will be made available at the CDS.
}} 

\author{I. Fuentes-Carrera \inst{1,2}
\and P. Jablonka \inst{3,4} 
\and A. Sarajedini \inst{5} 
\and T. Bridges \inst{6} 
\and G. Djorgovski \inst{7} 
\and G. Meylan \inst{4} }

\offprints{I. Fuentes-Carrera}

\institute{GEPI, Observatoire de Paris, CNRS, Universit\'e Paris Diderot, 
5 Place Jules Janssen, 92190, Meudon, France\\
\email{isaura.fuentes@obspm.fr}
\and
Instituto de Astronom\'\i a, Geof\'\i sica e Ciencias
  Atmosf\'ericas, Universidade de S\~ao Paulo, 
Rua do Mat\~ao 1226-Cidade Universit\'aria, 05508-900 S\~ao Paulo SP, Brazil\\
\and
Observatoire, Universit\'e de Gen\`eve
  chemin des Maillettes 51, CH-1290 Sauverny, Switzerland \\
\email{Pascale.Jablonka@obs.unige.ch}
\and
Laboratoire d'Astrophysique, Ecole Polytechnique F\'ed\'erale de Lausanne (EPFL),
Observatoire, CH-1290, Sauverny, Switzerland \\
\email{georges.meylan@epfl.ch}
\and
University of Florida, Department of Astronomy, 
Gainesville,FL 32611 USA \\
\email{ata@astro.ufl.edu}
\and
Department of Physics, Engineering Physics, and Astronomy, Queen's University, Kingston, ON K7L 3N6, Canada\\
\email{tjb@astro.queensu.ca}
\and
California Institute of Technology, Pasadena, CA 91125, USA\\
\email{george@astro.caltech.edu}
}

\date{Received . . . . . . . . . .  ; accepted . . . . . . . . . . }

\abstract
{}
{We quantify the intrinsic width of the red giant branches of three 
massive globular clusters in M31 in a search for metallicity spreads within these objects.}
{We present HST/ACS observations of three massive clusters in 
M31,  G78, G213, and  G280. A  thorough  description of the photometry
extraction and calibration is presented.  After derivation   of the
color-magnitude diagrams, we quantify the  intrinsic width of the  red
giant branch of each cluster.} 
{This width translates into a metallicity
dispersion that indicates  a  complex star formation  history  for this
type of system. For G78, $ \sigma_{[Fe/H]}=0.86 \pm 0.37 $; 
for G213, $ 0.89 \pm 0.20 $; and  for G280, $1.03 \pm 0.26$. 
We  find that the metallicity dispersion of the clusters does not scale with mean metallicity. We also find no trend with the cluster mass. 
We discuss some possible formation scenarios that would explain  our
results.}
{}


\authorrunning{Fuentes-Carrera et al}
\titlerunning{Metallicity spread in M31 globular clusters}

\maketitle

\section{Introduction}

There are now a few dozen Galactic globular clusters for which high
resolution spectroscopy  has allowed insight into their star-to-star
chemical
variations \citep{gratton2004}.  Omega Centauri ($\omega$ Cen) appears
to be the only case where such variations are  clear for all elements.
Several  scenarios  have  been  proposed  to  explain     these
observations:   The high  primordial  mass  of  $\omega$ Cen  might be
enough  to induce secondary  star formation  \citep{dopita86}.  Models
of chemical evolution have considered self enrichment scenarios
\citep{iku00,tsuji03}. Primordial chemical inhomogeneities have
also been investigated (e.g.,\cite{kraft94}). Finally, the possibility
that  $\omega$ Cen is the  nucleus  of a now-dissolved nucleated dwarf
galaxy    has  been   raised    by       a   number  of        studies
\citep{zin88,lee99,hilker2000}.

Our knowledge of extragalactic globular clusters  is currently far less
advanced.  At the moment,  the globular cluster G1 in  the halo of M31
constitutes the only evidence  for a metallicity spread, as derived from
the width of its red giant branch (RGB) \citep{meylan01}.  The parallel with
$\omega$  Cen is immediate.  Both  clusters  exhibit a metal abundance
range of about 1 dex, G1 being twice as  massive as $\omega$ Cen.  The
two clusters differ  in  a number of   ways though, which  are  likely
important clues to the understanding of their formation.  For example,
G1 and $\omega$   Cen have very  different locations  with respect  to
their  host galaxy - G1   is a halo   cluster located in projection at
$\sim$ 40 kpc from the center of  M31 (nearly the distance between our
Galaxy and the LMC),  while $\omega$ Cen is  now at $\sim$ 6 kpc  from
the Galactic center and at $\sim$ 1 kpc from the Galactic plane.

Although  $\omega$ Centauri is by  far the  brightest and most massive
globular  cluster in our Galaxy, G1  may not be  the only such massive
cluster belonging  to M31. There are  at least three other bright clusters
which  have central  velocity  dispersions larger than  20 km s$^{-1}$
\citep{djor97}.  They offer a unique   opportunity to investigate the
degree to which the  mass  of  a   globular  cluster influences   its
chemical
evolution.  In the velocity dispersion {\it versus} absolute magnitude
relation  (the equivalent for globular   clusters of the Faber-Jackson
relation),  the  two sequences of  Galactic  and  M31 systems show  no
detectable   difference  in    the  slope  and  zero-point   of  their
correlations.  Thanks to the similarity  between  the Galactic and M31
globular cluster  sequences, any conclusion  drawn from additional M31
clusters may represent general (universal) characteristics.

We present  here the analysis of images  obtained  with the  Advanced
Camera for Surveys  aboard the Hubble  Space Telescope of those  three
massive and bright clusters  in  M31.  Our  goal  is to quantify   the
intrinsic width of their red giant branches in order to search for any
metallicity spread.

\section{Observations and Data Reduction}

\subsection{Sample selection}

Velocity dispersions  are a good indicator  of dynamical  mass for old
stellar systems with nearly  constant mass-to-light (M/L) ratios, such
as  globular clusters \citep{djor94,djor97}.  Therefore, our selection
criterion for this work operates via  the cluster total luminosities and
velocity dispersions.

\begin{figure*}
\centering
\includegraphics[scale=0.6]{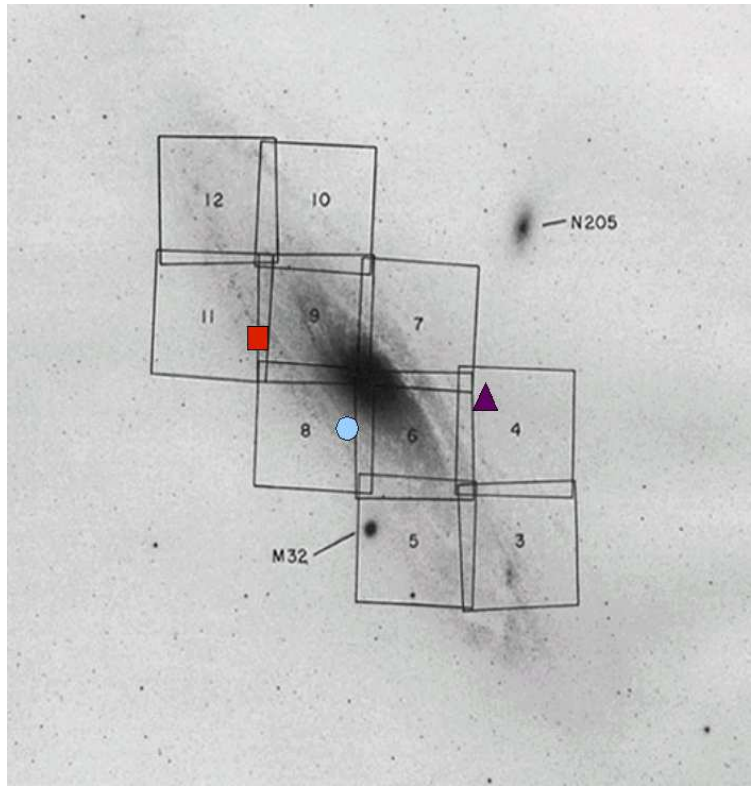}
\caption{Location of globular clusters in M31. Dark purple triangle corresponds to the location of G78, light blue circle corresponds to G213 and dark red square shows the location of G280. Numbered squared regions correspond to charts in the Atlas of the Andromeda Galaxy \citep{hodge81} }
\label{loc_clusters}
\end{figure*}

As mentioned previously, apart from G1,  three other clusters among the
21 objects in   the sample  of  \citet{djor97}  have central  velocity
dispersions,   $\sigma$,   greater than  20  $km   \  s^{-1}$.  Figure
\ref{loc_clusters} shows  the  location   of the  three  clusters   in
M31. All three of them  are located off the  major axis of M31 towards
the minor axis.  G78 is located (in projection) on the  edge of a dust
cloud  (object    D48 in  Hodge 1981)   which  lies  northwest  of the
cluster. G213  coincides in projection with  one of  M31's spiral arms
but   does  not lie    near   any particular dust   cloud   or stellar
association.  G280 is also located at the edge of a dust cloud (object
D642) lying to the northeast of the cluster.  Table
\ref{props_amas} lists some of the properties of these clusters.
Columns (1) and (2) give the coordinates of each cluster, column (3) the magnitude in the V filter \citep{huch91}, column (4) the projected distance from the center of M31, column (5) the metallicity \citep{huch91}, column (6) the velocity dispersion \citep{djor97} and column (7) the reddening value (Jablonka, Alloin \& Bica 1992 for  G78; this work for G213; Frogel, Persson \& Cohen for G280).

\begin{table*}
\centering
\caption{Properties of observed clusters.}
\label{props_amas}
\begin{tabular} {l c c c c c c c }
\hline\hline
Cluster & $\alpha$ (2000) & $\delta$ (2000)  & V(mag)  & $D_{M31}$(')
 & $ [Fe/H]$   &  $\sigma$(km/s) & $E(B-V)$  \\
         & (1)            &  (2)             &  (3)     &  (4)                 &  (5)        &  (6)             & (7)    \\
\hline
G78  & 00 41 01.2 & +41 13 45.5 & 14.2 & 19.2 & $-$0.92 & 25.48 & 0.23 \\
G213 & 00 43 14.4 & +41 07 20.5 & 14.5 & 11.2 & $-$1.08 & 20.50 & 0.10  \\
G280 & 00 44 29.5 & +41 21 35.8 & 14.3 & 20.5 & $-$0.70 & 25.94 & 0.10 \\ 
\hline
\end{tabular}
\begin{list}{}{}
\item[$^{\mathrm{ }}$] References: (3) and (5) Huchra, Brodie \& Kent 1991; (6) Djorgovski et al. 1997; (7) Jablonka, Alloin \& Bica 1992 for  G78, this work for G213, Frogel, Persson \& Cohen for G280.
\end{list}
\end{table*}

\subsection{Observations}

Observations of the  three clusters and  their surrounding fields were
obtained with the High Resolution Channel (HRC) of the Advanced Camera
for Surveys (ACS) aboard HST (Cycle 12; program ID 9719). The HRC has
a plate scale of 0.027$\arcsec \ pixel^{-1}$ and a field of view (FoV)
of $26\arcsec \times 29\arcsec$. Four  images of each globular cluster
were taken using the F606W (close to V) and F814W (close to I) filters
with total integration time of 505  x 4 = 2020 $s$  and 715 x 4 = 2860
$s$,  respectively.  In  order to be  able  to detect cosmic  rays and
other artifacts as  well as to sample  the PSF optimally, images  were
dithered    following  the  four-point    dither  pattern    known  as
ACS-HRC-DITHER-BOX with a step  of 0.15$\arcsec$ each\footnote{See the
ACS           Phase  II          Proposal      Instructions       {\it
http://www.stsci.edu/hst/acs/proposing/dither} for further details}.

\begin{figure*}
\centering
\includegraphics[scale=0.4,angle=0]{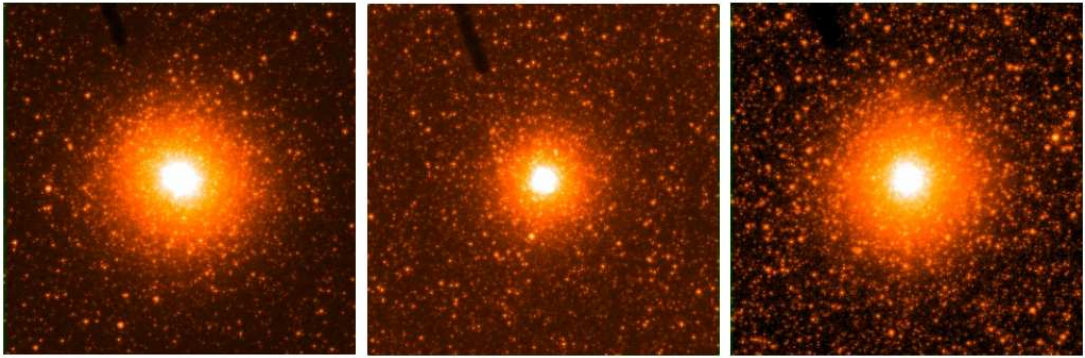}
\caption{{\it MultiDrizzled} images of G78 ({\it left}), G213 ({\it middle}) and G280 
({\it right}) and surrounding field stars in the F814W filter. North is to the top, 
East is to the left. The dark feature seen on the northeast corresponds to the occulting 
finger of the ACS. Field of view (FoV) for each cluster is 17.3$\arcsec \times$ 17.3 $\arcsec$.}
\label{imgs}
\end{figure*}

For each   cluster, we used  the  FLT images  downloaded  from the HST
archive. The instrumental signature has been removed from these images
(bias, dark  and flat-field  correction) by the on-the-fly-reprocessing
({\it OTFR})  performed by {\it CALACS}.  These  FLT images  were then
corrected  for geometric distortion    and  combined using the    {\it
MultiDrizzle}  task of    the STSci pipeline    incorporated  in PyRAF
\citep{koek02}.  This procedure eliminated all  of the cosmic rays and
bad pixels on the images.  We have used the default  values for all of
the parameters in the {\it MultiDrizzle}  task.  We used ``median'' as
the IMCOMBINE method  in order to  add up the  four images per filter.
Due to a gradient in the background of  the ACS field, images were not
sky-subtracted.   Background  variation   was dealt  with   during the
photometry    extraction  using  DAOPHOT     (Section \ref{phot}).  No
oversampling was  done when drizzling the  images into the final image
to avoid introducing a second redistribution of noise in the resampled
pixels.  Figure  \ref{imgs} shows the  final {\it MultiDrizzled} image
of each cluster  in the F814W filter.  The long, dark  feature on each
image is   the    shadow of  the   occulting finger   of   the ACS/HRC
coronagraph.  A quick  comparison of  these images  shows that G78  is
bigger than G213  and  G280  which are  rather  similar  in size   and
structure.  All of them have a  dense core, though proportionally, the
core  of G78 is  smaller with respect  to the cluster  as a whole. The
field surrounding  G213 is the   most  populated. Both  of the  fields
surrounding G78 and G280 are sparse, although that of  G78 seems to be
somewhat less populated.

\section{Photometry}
\label{phot}

\subsection{Photometry Extraction}
\label{phot_ext}

Photometry    was determined    using  the  DAOPHOT/ALLSTAR   routines
\citep{stet94}.    Since {\it  MultiDrizzled}  images  are produced in
units of  $electrons  \ second^{-1} \   pixel^{-1}$, they need   to be
multiplied by the exposure time and divided by  the gain factor to get
back  to the original  $ADU \  pixel^{-1}$ units  used  by DAOPHOT and
ALLSTAR.   In order to   do so we need   to consider how many original
images are  included in the  drizzled  images.  If  N images have been
combined by the  drizzle routine, then the read  noise is reduced by a
factor between 1 and $\sqrt N$, and the gain  is increased by a factor
between 1 and  N.   Unfortunately these  numbers are   not necessarily
constant because any one pixel in the  drizzled image can come from as
few as 1 or as  many as N original images.  One must also consider the
fact that the {\it MultiDrizzle} task redistributes  the noise in each
pixel of the image, while DAOPHOT assumes that the noise properties of
an image are approximately constant everywhere. In order to take these
effects into account in the   derivation of precision photometry,   we
followed  P. Stetson's suggestion and modified  our initial values for
the READ OUT  NOISE  and the GAIN   in both  the DAOPHOT  and  ALLSTAR
routines in order to have $\chi^2$ values close to 1. Object detection
was done   with     DAOFIND  on  the  combined     {\it MultiDrizzled}
image. Point-spread functions (PSFs) were constructed with DAOPHOT and
stars were measured using the ALLSTAR routine. The PSF of each cluster
on each filter  was built from bright isolated  stars in  the FoV. For
G213 and G280, the  PSF for the F606W image  was best determined using
a few, bright and  isolated stars (23  for G213 and  25 stars for  G280)
while for  the F814W image,  the  PSF was  better  computed using more
stars (52 for  G213 and 47  stars for G280).  For G78,  both PSFs were
best  determined using a  small  number of  stars  (22 stars  for both
images). For the three clusters, the stars used to estimate the PSF cover the whole FoV except for the central parts of the FoV (about 2 arcsec from the center of the FoV). 
As  a  first approach,  we used a  PSF  that  varied linearly
across the FoV in case there still remained a  slight variation of the
pixel scales due to the  optics of the  ACS.  Through the analysis  of
residuals after  star subtraction,  we  found that the  photometry was
best achieved using a spatially constant PSF,  which supports the fact
that the   pixel  variation   was properly   corrected   by   the {\it
MultiDrizzle} procedure.

\subsection{Photometric Calibration}

Once the instrumental photometry   was extracted, it was necessary  to
convert it to the STMAG system \citep{koor86}. In order  to do so, the
{\it MultiDrizzled} nature     of  the  images must   be   taken  into
account.  Our  {\it  MultiDrizzled}   images  were obtained   by  {\it
drizzling}  $N$ FLT images into a  single image. FLT images have units
of $electrons$  while the {\it  MultiDrizzled} image  is  in units  of
$electrons \ \ s^{-1}$. Photometry was extracted using the DAOPHOT and
ALLSTAR  packages which work with $ADU$.   As previously mentioned, in
order to work with  a {\it MultiDrizzled}  image using these packages,
we multiplied  the original {\it  MultiDrizzled} image by the exposure
time of one FLT  image, $t_{exp}$, and  divided it by  the gain of the
CCD, $gain_{ccd}$\footnote{ This  can only be  done if  ALL FLT images
used in  the {\it drizzling}   process have  the same  exposure  time,
$t_{exp}$}.  We shall call  the  resulting  image, the {\it   modified
Multidrizzled} image.

\subsubsection{Charge Efficiency Transfer corrections}

Photometric  losses due to  the problem  of charge transfer efficiency
(CTE) of the CCD need  to be quantified before converting instrumental
magnitudes to any photometric   standard system. For the  ACS/HRC, the
CTE   correction  is   given  by   the    formula  in   the ACS   Data
Handbook\footnote{chapter 6,   section  1.5}.  In   order to have  the
correct  input values, the  SKY brightness obtained  with ALLSTAR from
the   {\it  modified MultiDrizzled}  image    in  {\it  ADU}, must  be
multiplied by the gain  of the CCD. On  the other hand, the $FLUX$  of
the star (in $electrons$) is the original  flux on the FLT image. This
flux is  equal  to  the flux  associated   to the magnitude given   by
DAOPHOT,   $MAG_{modDRZ}(star)$, times the   gain     of the CCD,    $
gain_{ccd}$. Considering that DAOPHOT adds a  zeropoint of 25.0 to all
magnitudes, the resulting flux is given by

$$ FLUX \ (electrons) = 10^{-(MAG_{modDRZ}(star) - 25.0)/2.5} \times
{ gain_{ccd}}  \eqno(1)$$

Once the $SKY$  and $FLUX$ values  have been corrected, the  resulting
CTE correction ($YCTE$) will be that for a single FLT image.

\subsubsection {Aperture Corrections}

Our aim is to transform the magnitudes obtained from ALLSTAR, which we
will call {\it PSF  magnitudes}, into magnitudes  in the STMAG system.
In  order to do   so, several aperture  corrections need  to  be made.
Before  being converted  into    magnitudes   in an   aperture    with
``infinite''  radius, {\it  PSF magnitudes}  need  to be corrected  to
magnitudes in an  aperture of radius 0.5" (18.5  HRC $pixels$).  Since
the aperture   corrections used  to  transform instrumental magnitudes
into the STMAG  system have been  computed  for single FLT images  (in
units of $electrons$), we need to  ``divide'' all {\it PSF magnitudes}
derived from      the   {\it      modified   MultiDrizzled}     image,
$MAG_{modDRZ}(PSF)$ in $ADU$, by the gain of the CCD.

$$ MAG_{FLT}(PSF) = MAG_{modDRZ}(PSF) \  -  \  2.5 \ \log(gain_{ccd})    \eqno(2)$$

We then select 10-20 uncrowded and bright  (F606W and F814W magnitudes
brighter than  16.5 mag and 15.5 mag,  respectively) stars on the {\it
MultiDrizzled}  image  and measure   their total magnitudes   inside a
radius of $18.5     \   pix$, $MAG_{modDRZ}(18.5)$,  using    PHOT  in
DAOPHOT. Since this magnitude value will be  that corresponding to the
{\it modified MultiDrizzle} image (whose flux is in $ADU$), we need to
convert  it into   magnitudes on the  FLT  image   (whose  flux is  in
$electrons$).

$$  MAG_{FLT}(18.5) = MAG_{modDRZ}(18.5)  \ - \ 2.5 \ \log(gain_{ccd})
\eqno(3)$$

For each image, we calculate the mean  value of the difference between
these  two magnitudes.  The   mean value  of  this  difference is  the
aperture correction, $AP\_CORR(18.5)$, for each FLT image.  This value
is   then added to   all  of the  PSF   magnitudes  of the FLT  image,
$MAG_{FLT}(PSF)$.  The  resulting magnitudes, $MAG_{apcorr}$,  need to
be  converted from  this ``intermediate'' aperture  to ``infinity'' or
{\it  observed magnitude},  $OBMAG_{FLTinf}$.   This   is done   using
equation (2)  in  \citet{sir05}.  Strictly  speaking, the Sirianni  et
al.  corrections  to infinity   are only  applicable to   the drizzled
images, not   the FLT images.    Nevertheless,   in a  recent   paper,
\citet{ata06} show   that there is  no significant  difference between
them.  In order to  use this equation, $MAG_{apcorr}$  needs to be set
to  $COUNT\_RATE$ in   $electrons  \  s^{-1}$  through the   following
equation

$$ COUNT\_RATE (electrons \ s^{-1}) = {{10^{MAG_{apcorr}(18.5)/2.5}} \over t_{exp}}  \eqno(4)$$

Once all  magnitudes have  been  set to  the {\it observed magnitude},
$OBMAG_{FLTinf}$, they  need  to be corrected  by the   CTE correction
previously  computed.  The $ OBMAG_{final}$ will  be used to transform
the instrumental  magnitude   system  to the  STMAG  system  following
equation (7) of \citet{sir05}.

\subsection{Conversion to STMAG absolute magnitudes}
\label{conv_absmag}

Finally the  observed STMAG  magnitudes   need to be  converted   into
absolute STMAG magnitudes  taking into account  the distance at  which
the clusters lie  and their reddening  value. This was  done following
the   procedure  by  \citet{brow05}. A   value   of $(m-M)_0 =  24.43$
\citep{freed90} was  used   for the  true  distance  modulus. For each
cluster, two reddening values were considered:  0.1 and 0.23 The value
$E(B-V)$=0.1 is the Milky  Way line of  sight reddening in the direction
of   M31.      It  is     the   value     estimated    for   G280   by
\citet{frog80}.  $E(B-V)$=0.23  is   the  value estimated   for  G78  by
\citet{pj92}. There is no reddening value in the literature for G213.

\section{Statistical Selection of Stars}
\label{stat_sel}

\begin{figure*}
\centering
\includegraphics[scale=0.57,angle=270]{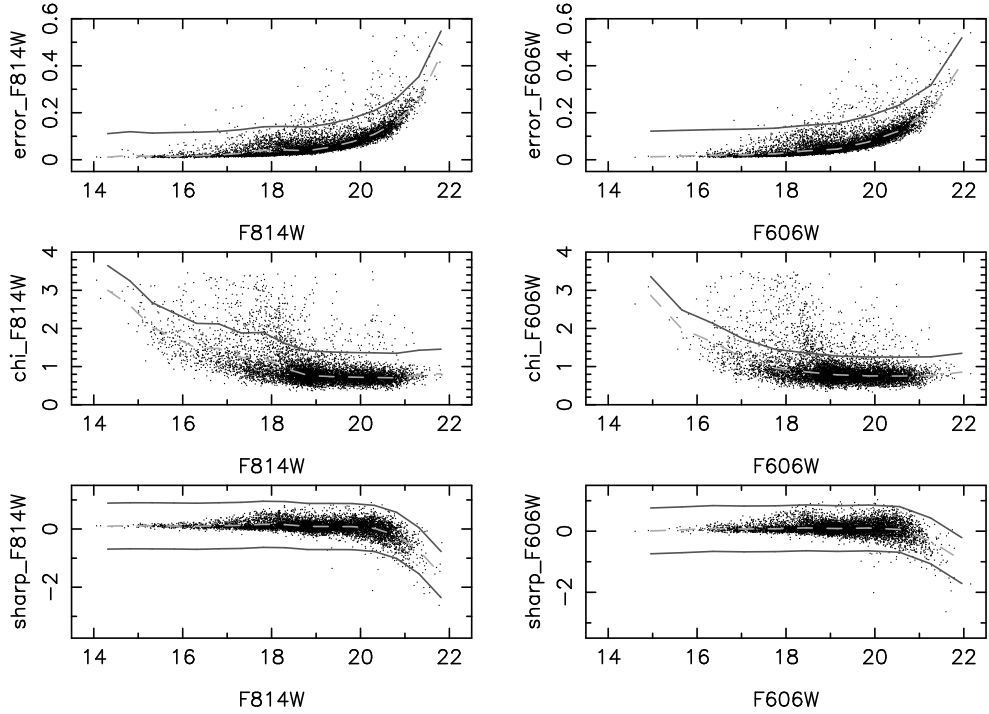}
\caption{  DAOPHOT/ALLSTAR output for G78.
{\it Top:} Error vs. magnitude for both F606W and F814W instrumental magnitudes. 
{\it Middle:}$\chi^2$ vs. magnitude. {\it Bottom:} Sharpness vs. magnitude. 
Solid lines show the upper limits for stars used to build the CMDs of the cluster 
and the field; dashed line shows the averaged value after iteration. In the case of the 
sharpness a lower limit was also used to select stars.}
\label{G078_selection}
\end{figure*}

\begin{figure*}
\centering
\includegraphics[scale=0.57,angle=270]{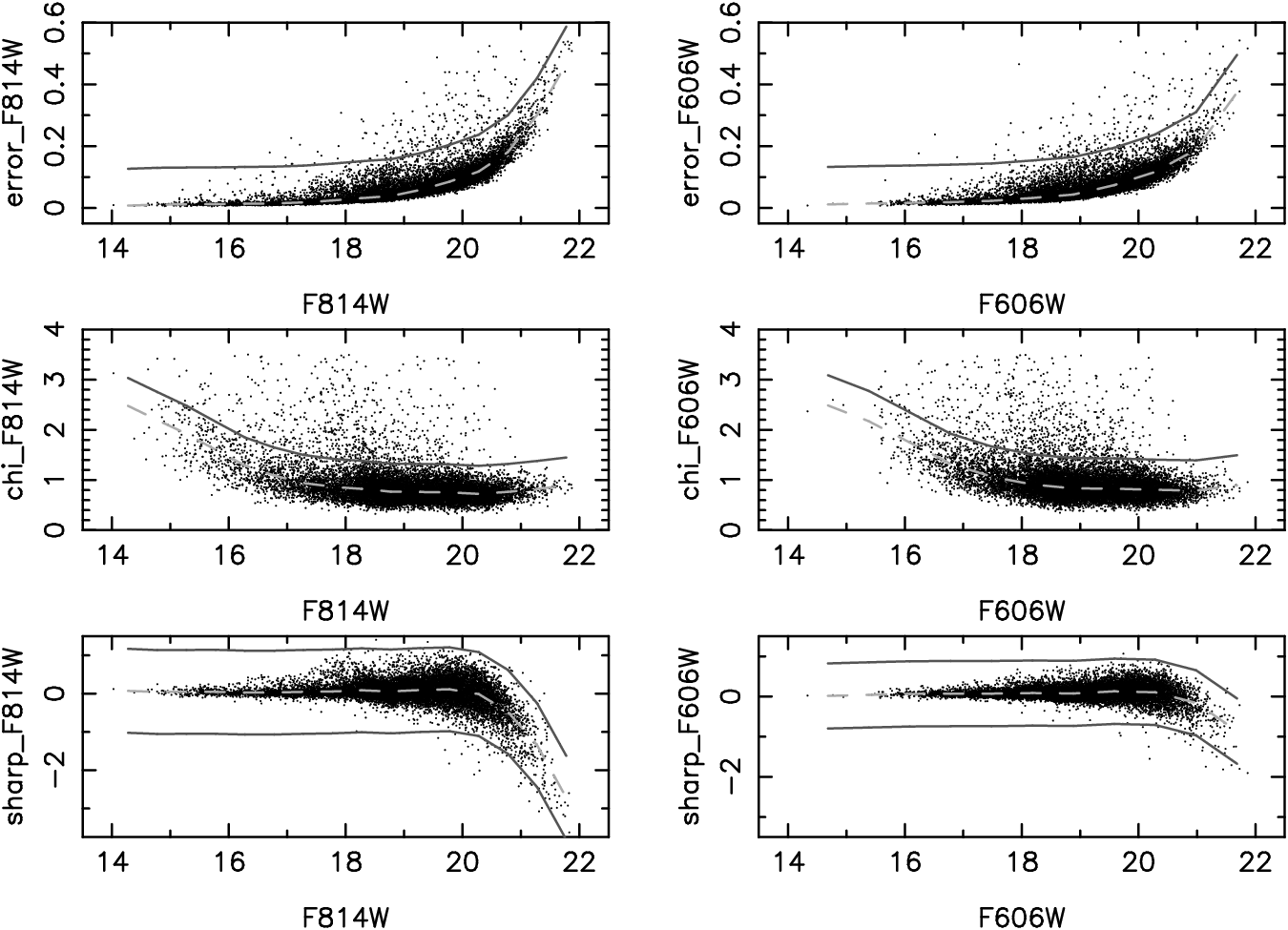}
\caption{  DAOPHOT/ALLSTAR output for G213. Same as Figure \ref{G078_selection}.}
\label{G213_selection}
\end{figure*}

\begin{figure*}
\centering
\includegraphics[scale=0.57,angle=270]{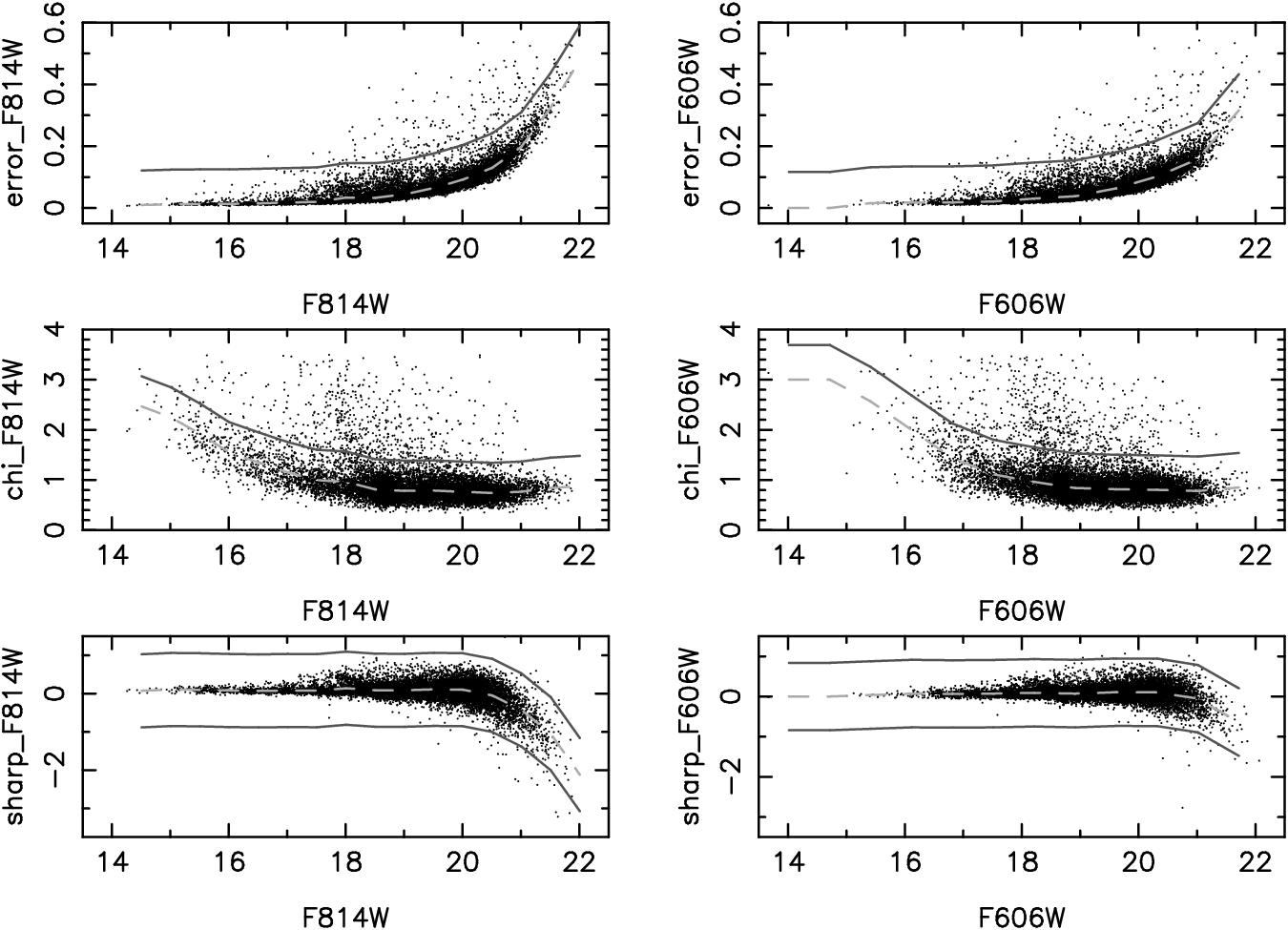}
\caption{  DAOPHOT/ALLSTAR output for G280. Same as Figure \ref{G078_selection}.}
\label{G280_selection}
\end{figure*}

Since we  are interested in determining  the width of  the RGB of each
cluster, we need to eliminate any spurious photometric detections that
might     contaminate our results.  In order     to  do so, stars were
statistically selected  to eliminate  objects with dubious  values for
the  error,  $\chi^2$ and   sharpness.  This  was done   prior  to the
photometric calibration,  that is with  the  star magnitudes given  in
instrumental           magnitudes.Figures        \ref{G078_selection},
\ref{G213_selection}   and \ref{G280_selection}   show  the output  of
DAOPHOT for the total ACS field of  the three clusters.  For all three
parameters, most points  seem to follow a  global trend. For instance,
in the case  of G280, error values increase  from $\sim$ 0.02 at  16.0
F814W to 0.4 at 22.0 F814W. There is however, a percentage of outliers
for  each distribution for  all magnitude  values. We considered these
points   as possible  spurious photometric   detections to be  removed
before any analysis. A statistical  selection was done on the estimate
of the dispersion  and  the mean  value  of the  distribution of  each
parameter.  The  mean  value    of   the distribution   was   computed
recursively;  after  each  computation, stars   beyond  3 sigma of the
distribution  were eliminated before the     next computation of   the
mean. A bin of 0.7 mag was used for parameters in the F606W filter and
a bin of 0.5 mag in the F814W filter.  The computation of the mean was
performed until the difference  between subsequent values of  the mean
value of each parameter became lower than 1\% of  the value. Once this
final  mean value was determined, we   considered the highest value of
the dispersion of  each  parameter  (since  the computation  was  done
independently for each magnitude  bin the dispersion values vary  from
one bin to the other). This value for the dispersion was multiplied by
a factor  between 1.0 and  3.5 in  order to  delimit the edges  of the
distribution. These  factors  were   chosen  in  order  to  draw   the
boundaries that follow the bulk  of the distribution of each parameter
value.      The   solid   lines    in  Figures   \ref{G078_selection},
\ref{G213_selection} and \ref{G280_selection} show the upper limit for
the statistical selection of stars; the  lower limit for the selection
was also considered in  the case of   the sharpness. The dashed  lines
show the average   value of each parameter.  For  all clusters, stars
whose $\chi^2$ values  were over 3.5  for both filters  were discarded
before any statistical selection was  done, in order to facilitate the
statistical selection and to eliminate stars that had undetermined sky
flux to be used for the CTE correction.

Following   this selection, we  were left   with  7770 stars  from the
initial 8139 stars for G78.  For G213, 13275  stars from 14781 initial
stars were selected. For G280, 11204 stars from 12409 were kept.

\section{Artificial Stars Experiments}
\label{ASE}

\begin{figure}
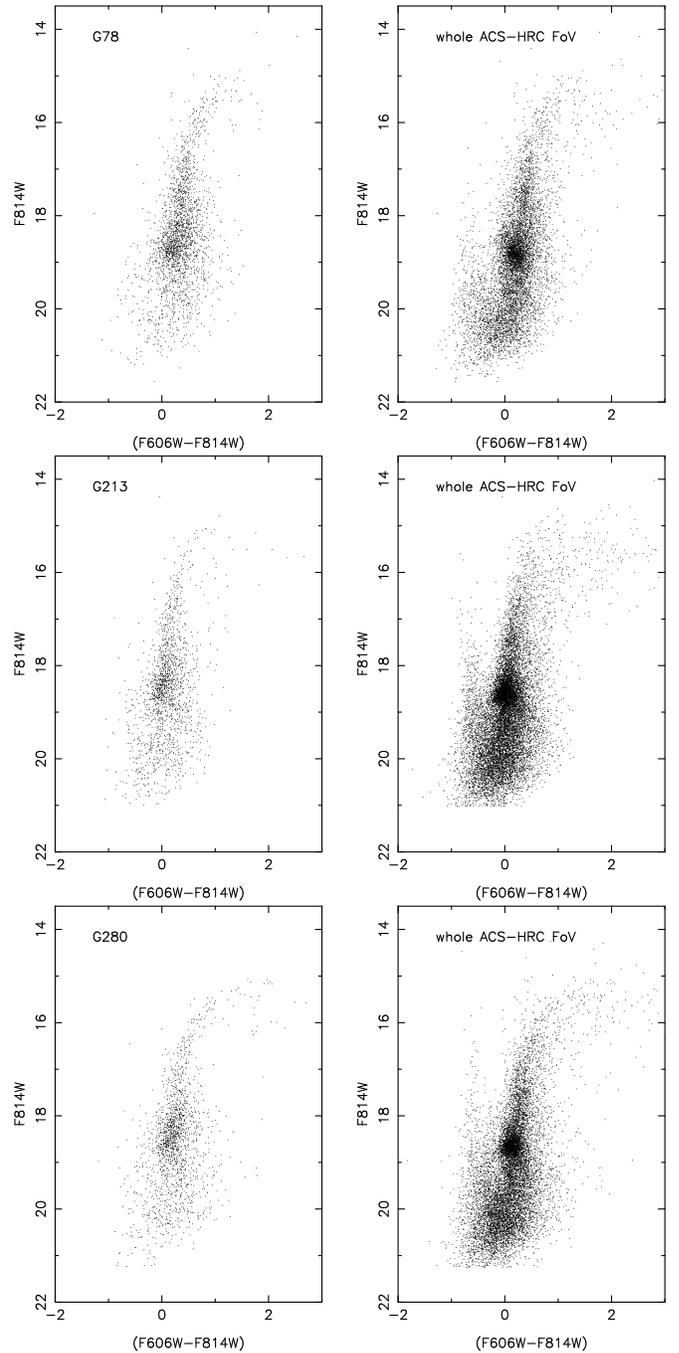

\centering
\includegraphics[scale=0.38,angle=270]{G078_CMD_amas-vs-FoV_jan08.ps}
\includegraphics[scale=0.38,angle=270]{G213_CMD_amas-vs-FoV_fev08.ps}
\includegraphics[scale=0.38,angle=270]{G280_CMD_amas-vs-FoV_fev08.ps}
\caption{{\it Left panels}: Color-magnitude diagrams (CMD) for G78 ({\it top}), G213 ({\it middle}) and G280 ({\it bottom}) considering all stars within $R_{out}$ -values given in Table \ref{avg_disp_ase}. {\it Right panels}: CMDs considering all the stars in the field of view (FoV) of the {\it MultiDrizzled} HST-ACS images of G78 ({\it top}), G213 ({\it middle}) and G280 ({\it bottom}).}
\label{CMDs_inst}
\end{figure}

\begin{figure}
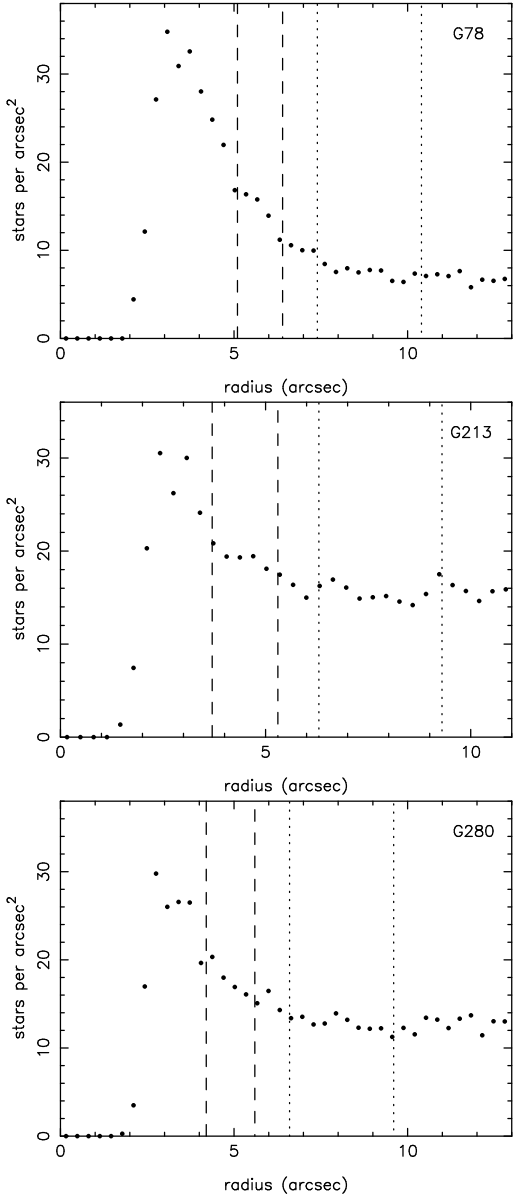

\centering
\includegraphics[scale=0.45,angle=270]{G078_dens_jan08.ps}
\includegraphics[scale=0.45,angle=270]{G213_dens_fev08.ps}
\includegraphics[scale=0.45,angle=270]{G280_dens_fev08.ps}
\caption{Density profile for all stars in the FoV of G78 ({\it top}), 
G213 ({\it middle}) 
and G280 ({\it bottom}) after statistical selection (see section \ref{stat_sel}). Dashed lines 
delimit the region used to construct the CMD of each cluster. Dot-dashed lines delimit the 
region of the FoV used for the CMD of the surrounding stars  
(see section \ref{cluster_vs_field}). Apparent density holes near the center of the clusters are 
due to star crowding.}
\label{dens}
\end{figure}

To evaluate  the photometry extraction and  to quantify the effects of
stellar crowding, we performed ten sets of artificial star experiments
per  cluster. These were done  using instrumental magnitudes. In order
to  do  so, we constructed the  CMD  of each  cluster  in instrumental
magnitudes. Left panels of Figure \ref{CMDs_inst} show the CMD derived
for each cluster, right panels show  the CMD built considering all the
stars in the  ACS FoV  -after  statistical selection. For the  cluster
CMDs, stars were selected within a circle  centered on each cluster.
This was done through the analysis of the radial distribution of stars
on  each FoV.  Figure \ref{dens}  shows the radial density profile for
the entire FoV centered on each cluster.
The  outer radius of the   cluster, $R_{out}$, was   chosen to be  the
radius  at which   the  density profile  dropped significantly  before
reaching a fairly constant  value  in order to minimize  contamination
from field stars. This plateau is assumed  to represent the density of
the surrounding field.  Values  for  $R_{out}$  are  shown  in  Table
\ref{sigmas}. We then  derived the fiducial RGB for each cluster  by
fitting the mean locii of stars between 15.8 and 18.2 mag in the
 F814W filter.  The faintest  limit of the magnitude interval was chosen  to
avoid the mixed stellar population of the red clump.

For each  experiment, we selected 384  stars along the  fiducial RGB of
each cluster from 15.8 to 18.2 in F814W in magnitude bins of 0.05 to
fully cover  the RGB interval. This resulted  in 8 stars per magnitude
bin per experiment and 80 stars per  magnitude bin  for   the ten
experiments as a whole - in total, 9216 artificial stars were used for
the experiments. The artificial stars were added to both the F606W and
F814W HRC/ACS {\it MultiDrizzled}  frames with the constraint that  no
two artificial  stars be  within  1.5  PSF  radii  of each  other  and
avoiding the  edges of  the images.   The number of   artificial stars
considered  per  experiment was  chosen to  be  small enough  to avoid
introducing important density variations  in the FoV of  the clusters.
The number of artificial stars introduced in the FoV is at most 5\% of
the total number of real stars in the field.   The photometry from the
resultant images  was computed following the  exact same steps used to
derive the   photometry     of  the observed    images   (see  section
\ref{phot_ext}).  The   percentage of  recovered   stars  for  all ten
experiments per cluster  is 97.94 \%, 98.13  \% and 98.29 \%  for  G78,
G213 and G280, respectively. The missing fraction are artificial stars
whose photometry could not be extracted with  DAOPHOT because they are
located too close ($< 3px$) to real stars in the FoV.

\begin{figure}
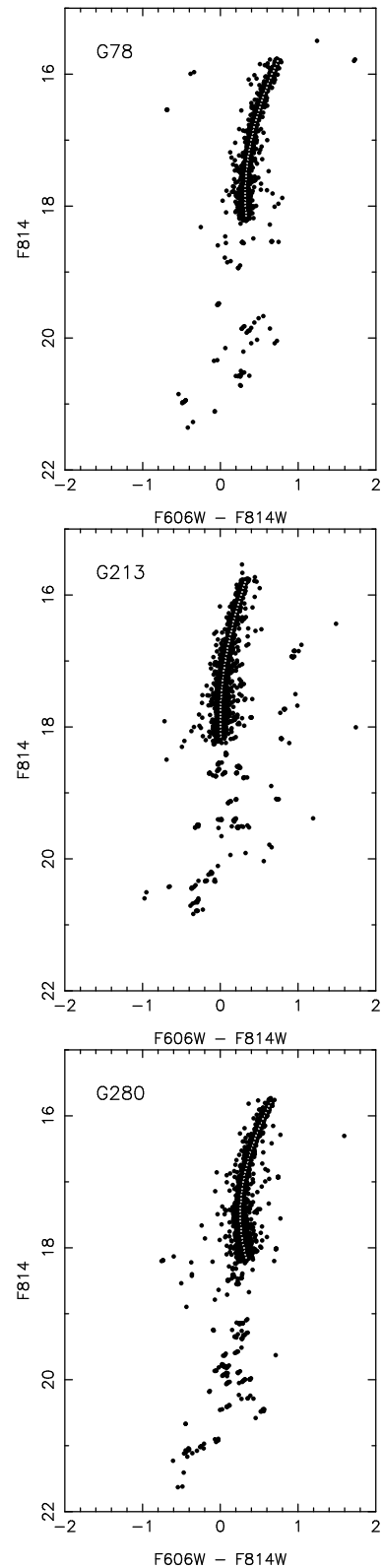

\centering
\includegraphics[scale=0.45,angle=270]{G078_CMD_ase-nosel-10exp_jan08.ps}
\includegraphics[scale=0.45,angle=270]{G213_CMD_ase-nosel-10exp_fev08.ps}
\includegraphics[scale=0.45,angle=270]{G280_CMD_ase-nosel-10exp_fev08.ps}
\caption{Results of the artificial stars experiments with no statistical selection of stars 
for G78 ({\it top left}), G213 ({\it top right}) and G280 ({\it bottom}). The thin white solid line indicates the 
input RGB while the dots indicate the output (measured)  RGB stars. }
\label{M31_ASE_nostat}
\end{figure}

Figure  \ref{M31_ASE_nostat} displays the   input artificial stars  along the
fiducial RGB and the  recovered RGB stars  for all the experiments  in
each cluster. No  statistical selection of  output stars was done  for
this   comparison. 
For  the three clusters,  the  input sequence is well recovered. 
Less than 5.0 \% of recovered stars fall  below the magnitude 18.2 in F814W -for G78 it is less than 2.5 \%. For this same cluster, less than 1.0 \% of the recovered stars  fall to the right (and/or left) of the fiducial RGB - between 1 and 2 magnitudes in color. The same occurs for the recovered stars of G280. For G213, there is a higher percentage  of stars  ($\sim  3.0 \%$)  falling outside the ``main'' body   of the RGB than  for  G78 and   G280.
This is probably due to the  fact that the  FoV of this cluster
is more populated   than those of G78  and  G280.  In all  cases,  the
number of outliers from the  recovered RGB is small. These  are
stars whose recovered magnitudes differ  significantly from the  input
magnitudes  because they happen to overlap with the original
(real) stars in the FoV.
Although  the artificial  stars were  placed  so  that no  two
artificial stars   are within 1.5  PSF radii  of each  other,  no such
restriction was imposed for stars already in the FoV.

\begin{figure}
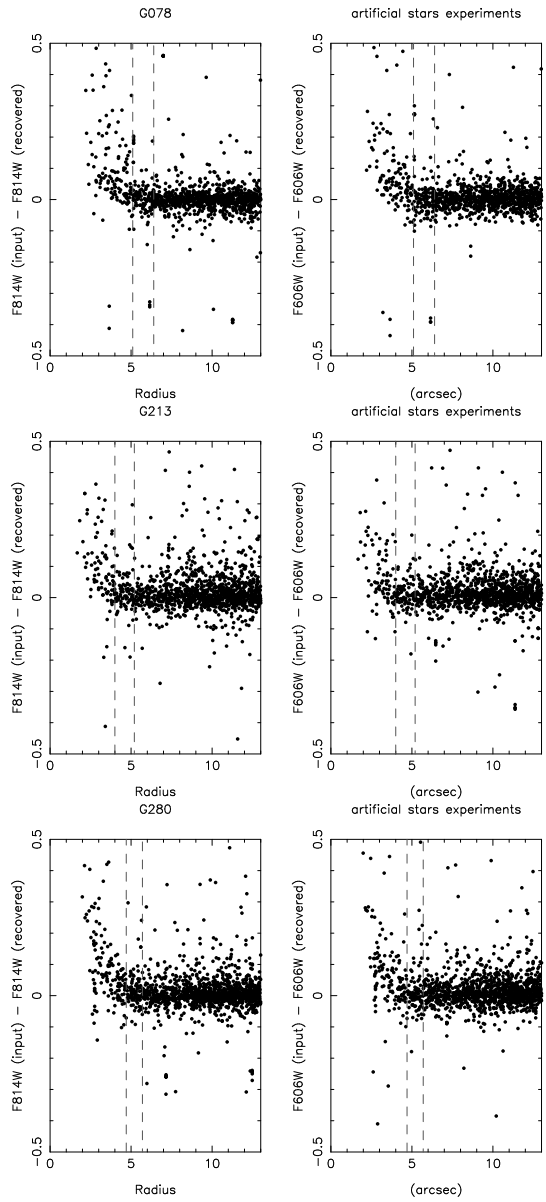

\centering
\includegraphics[scale=0.35,angle=270]{G078_ase_diff-rad-nosel-10exp_jan08.ps}
\includegraphics[scale=0.35,angle=270]{G213_ase_diff-rad-nosel-10exp_fev08.ps}
\includegraphics[scale=0.35,angle=270]{G280_ase_diff-rad-nosel-10exp_fev08.ps}
\caption{Input and recovered magnitude difference according to radial position for G78 ({\it top}), G213 ({\it middle}) and G280 ({\it bottom}). 
 For each cluster, the inner lines correspond to 5.1$\arcsec$, 3.7$\arcsec$ and 4.2$\arcsec$   for G78, G213 and G280, respectively, while the outer lines correpond to 6.4$\arcsec$, 5.3$\arcsec$ and 5.6$\arcsec$  for G78, G213 and G280,respectively.}
\label{ASE_mag_vs_rad}
\end{figure}

An evaluation of the   input minus  the  recovered magnitudes  in the   F606W and F814W
filters according to the position  of the star from  the center of the
cluster showed that this difference increases at small radii ($ R\sim 4.8\arcsec, 3.5\arcsec$ and $4.0\arcsec$  for G78, G213 and G280, respectively) 
introducing an artificial brightening for stars near the
center of each cluster.
Figure \ref{ASE_mag_vs_rad} shows the difference between the input magnitudes of the artificial stars and their recovered magnitudes ($mag\_diff$) according to their radial position with respect to the center of each cluster. An important departure from zero is seen for low values of R. For higher values, some stars do ``migrate" in magnitude, but they are few if compared to the bulk of the stars which fall around zero. 
In order to evaluate this brightening for the stars in each cluster, we computed the average value of $mag\_diff$ and its dispersion within $R_{out}$. 
We considered all stars within an annulus of outer radius $R_{out}$ and 
whose inner radius we varied until the absolute average value of $mag\_diff$ was less than 0.02 mag and the dispersion value was close to 0.05 mag or less. Results are shown in Table \ref{avg_disp_ase}.
This final radius was set to be the inner radius $R_{in}$ of the cluster. All stars lying within this radius will not be taken into account in the analysis that follows.

Sources of error, e.g., such as residual flat-field
non-uniformities and residual dark current, are not included in our 
photometric error
estimates. However, those should not  dramatically increase our estimates.
There is also a possibility that small
differences exist between the PSF model used in the artificial star
experiments and the profiles of the real stars. This source of error 
is difficult to quantify; however, we believe it is likely to be negligible
because of the spatial resolution of the ACS/HRC instrument which provides
good sampling of the PSF (FWHM $\sim$ 3 pixels).


\begin{table}
\caption{Average value and dispersion of the difference between the input magnitude and recovered magnitude between specific radial intervals for each cluster}
\label{avg_disp_ase}
\centering
\begin{tabular} {l c c c c}
\hline\hline
Cluster & radial interval ($\arcsec$) & mean magnitude difference & dispersion    \\
\hline
G78     & 5.1 - 6.4 & -0.0203 & 0.0454       \\
G213    & 3.7 - 5.3 & 0.0343  & 0.0551       \\
G280    & 4.2 - 5.6 & 0.0425  & 0.0528       \\
\hline
\end{tabular}
\end{table}

\section{Luminosity function}
\label{LF}

\begin{figure}
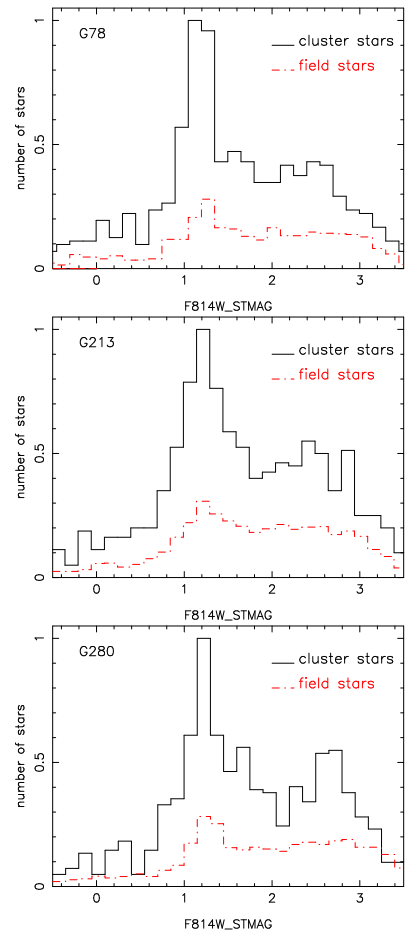

\centering
\includegraphics[scale=0.35,angle=270]{G078_lumfunc-e022_jan08.ps}
\includegraphics[scale=0.35,angle=270]{G213_lumfunc-e01_fev08.ps}
\includegraphics[scale=0.35,angle=270]{G280_lumfunc-e01_fev08.ps}\
\caption{Luminosity function (LF) for G78 ({\it top}), G213 ({\it middle}) and G280 ({\it bottom})
and their corresponding surrounding fields. 
 For each cluster, the LF was derived 
considering the following 
inner radii, $R_{in}=$5.1$\arcsec$, 3.7$\arcsec$, 4.2 $\arcsec$ for G78, G213 and G280 respectively. }
\label{M31_lf}
\end{figure}

\begin{figure}
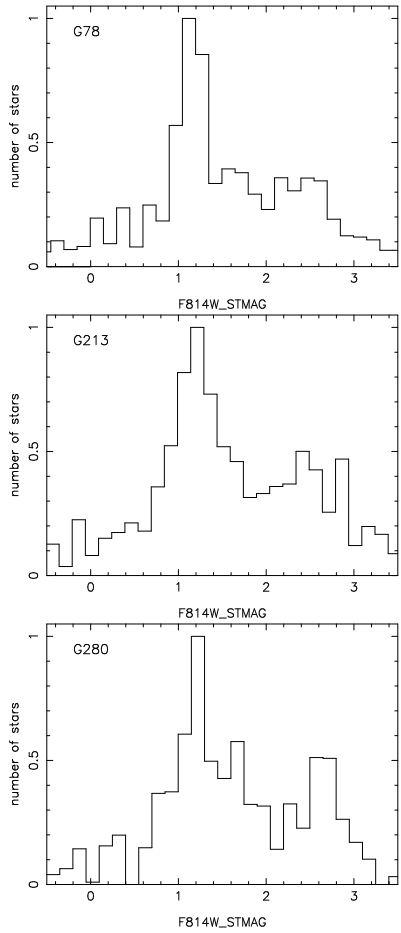

\centering
\includegraphics[scale=0.35,angle=270]{G078_LFsub-e022_jan08.ps}
\includegraphics[scale=0.35,angle=270]{G213_LFsub-e01_fev08.ps}
\includegraphics[scale=0.35,angle=270]{G280_LFsub-e01_fev08.ps}\
\caption{Luminosity function (LF) for G78 ({\it top}), G213 ({\it middle}) and G280 ({\it bottom}) subtracted from the contribution of their corresponding surrounding fields. 
}
\label{M31_lfsub}
\end{figure}

The  luminosity function (LF) for  each   cluster and its  surrounding
field   stars    was   derived     using    a   bin    of   0.15    in
magnitude\footnote{Absolute  STMAG magnitudes are considered from this
part of the  analysis onwards.}.  Field stars  were selected  within a
concentric annulus with an  inner radius 1.0''  larger than the  outer
radius  of the cluster, $R_{out}$ -defined  in section \ref{ASE}. This
was done to   reduce  the number  of   cluster  stars that   could  be
contaminating  the field.   The outer  radius of  the ring  containing
field stars was chosen 3.0'' larger than its inner radius.  
The LF of each cluster  was computed considering  all of the
stars between the inner radius, $R_{in}$ and outer radius, $R_{out}$ defined in the previous section.
Figure  \ref{M31_lf}  shows the
luminosity function for each cluster and its corresponding surrounding
field. Figure \ref{M31_lfsub}  shows the resulting luminosity function
of each cluster  subtracted from the  contribution of the stars in the
surrounding   field.   For G78,  an    $E(B-V)$ of  0.23 was  considered
initially  (taken from \citet{pj92}).  This  value was then changed to
0.22 in order to match the location (in magnitude) of the HB of G78 to
the location (in magnitude)  of the HBs  of G213  and G280.  For G280,
$E(B-V)$=0.1 \citep{frog80}.  For G213, there are no reddening values in
the literature, so  the  luminosity function was computed  considering
$E(B-V)$=0.1, which is taken to be the standard value along the M31 line
of sight. The  superposition of the CMD of  G213 with the standard RGB
of sequences of \citet{brow05}  confirms the validity of this  choice (middle panel of Figure \ref{ridges}). For each  cluster and its  surrounding  field, the  red clump  lies at
$\sim$1.185.

\section{RGB width determination}

\subsection{Final CMD}
\label{final_cmd}

\begin{figure}
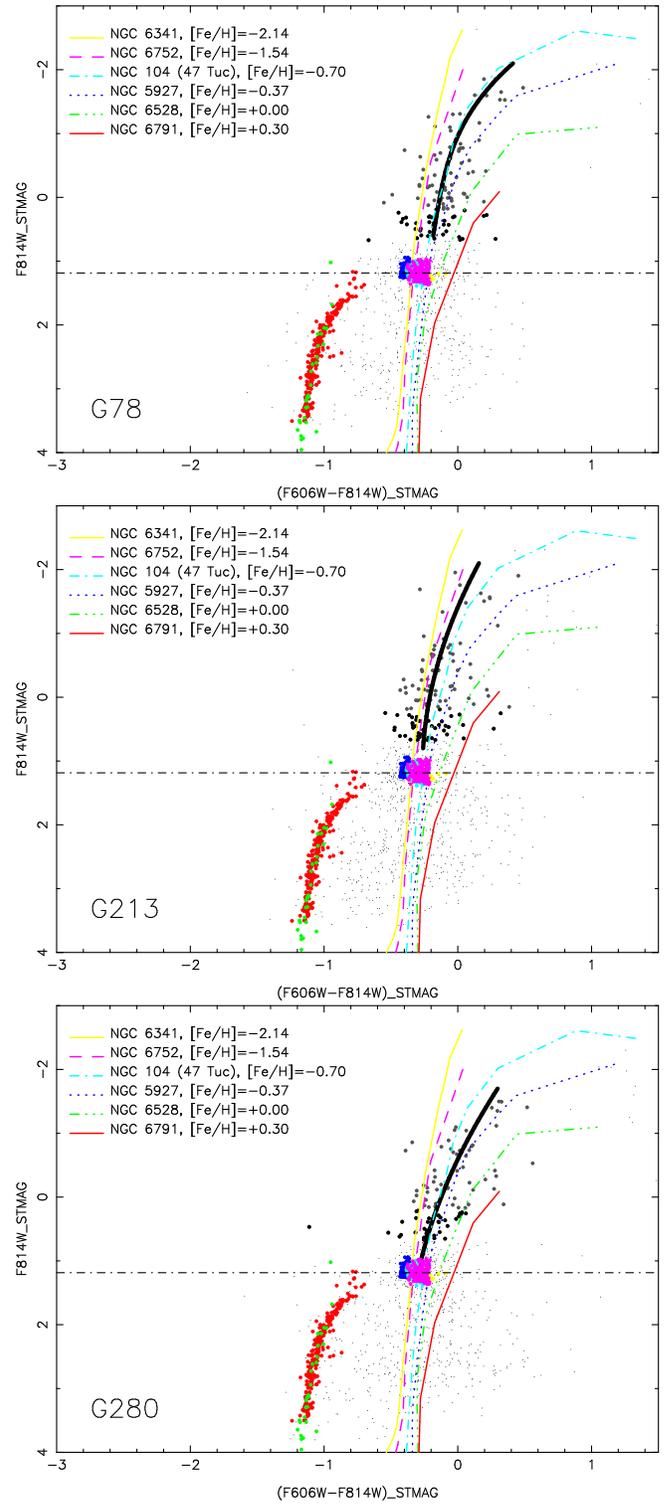

\centering
\includegraphics[scale=0.375,angle=270]{G078_ridges-e022_jan08.ps}
\includegraphics[scale=0.375,angle=270]{G213_ridges-e01_fev08.ps}
\includegraphics[scale=0.375,angle=270]{G280_ridges-e01_fev08.ps}
\caption{Superposition of the ridges and HB loci derived by \citet{brow05} on the CMD of  G78 ({\it top}), G213  ({\it middle}) and G280 ({\it bottom}).  Thick solid line indicates the position of the fiducial RGB of each cluster. The straight dotted line shows the position of the red clump of each cluster.  }
\label{ridges}
\end{figure}

The final CMD of each cluster was built considering only statistically
selected stars within the ring between $R_{in}$ and $R_{out}$.  Figure
\ref{ridges} shows the superposition of the standard RGB sequences and
HB  loci derived  by  \citet{brow05} on   the CMD   of G78, G213   and
G280. The straight dotted line shows the position of  the red clump of
each cluster. For the three clusters, the  red clump falls at the same
position as the HB loci of \citet{brow05}.

\begin{figure}
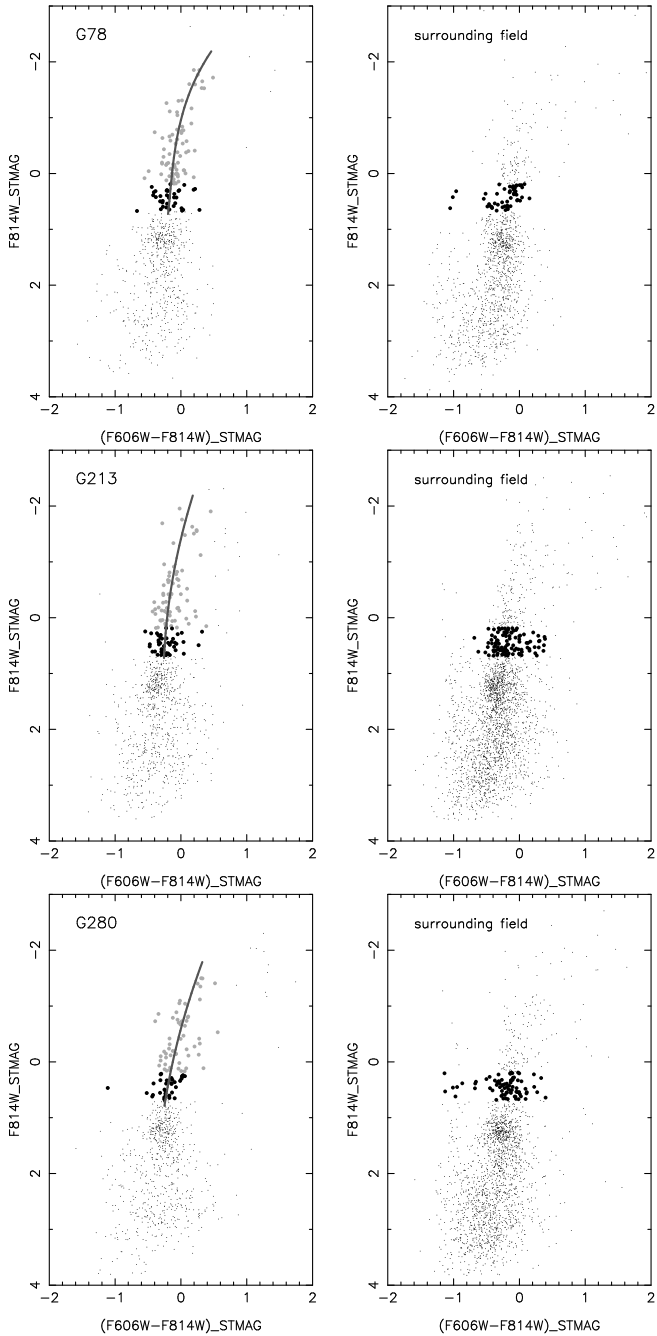

\centering
\includegraphics[scale=0.375,angle=270]{G078_CMD-e022_jan08.ps}
\includegraphics[scale=0.375,angle=270]{G213_CMD-e01_fev08.ps}
\includegraphics[scale=0.375,angle=270]{G280_CMD-e01_fev08.ps}
\caption{Color-magnitude diagrams (CMD) for G78 ({\it top left}), G213 ({\it middle left}) and G280 ({\it bottom left}) and corresponding surrounding fields, {\it top right} for G78, {\it middle right} for G213 and {\it bottom right} for G280. 
Grey dots indicate the stars considered for the fit of the RGB. Solid lines show the locus of this fiducial RGB. Black dots indicate the stars considered for computation of the RGB width. Right panels show the CMD for each surrounding region. Dark dots indicate stars considered for the analysis of the contribution of field stars to the width of the RGB of the cluster.}
\label{CMDs}
\end{figure}

At this point,  the  CMD of  each  cluster should  be  largely free of
spurious  detections. Nevertheless it   will still be affected by  the
presence of field stars  falling within the area  of each  cluster. In
order to estimate the  contribution of field  stars to the RGB of each
cluster, we also constructed the  CMD of the  stars in the surrounding
fields.  These stars were chosen within the outer annulii described in
section \ref{LF}. The left  panels of figure  \ref{CMDs} show the CMDs
of G78, G213 and G280.  Right panels of the  same figures show the CMD
of their corresponding surrounding fields.

For  the final CMD of each  cluster, the  fiducial  RGB was derived by
fitting  the mean locii  for stars within  certain magnitude and color
intervals. The upper magnitude limit was  set at 0.5 mag brighter than
the position of  the red clump. The  lower magnitude limit, as  well as
the color limits, were set in order to get rid  of obvious outliers of
the RGB. Magnitude  and color intervals for  each cluster are shown in
Table \ref{mag_col_int}. The fit is shown  in Figures \ref{ridges} and
\ref{CMDs}. The dark lines in the left  panels of these Figures show
the mean locii of the RGB regions of each cluster.

\begin{table}
\caption{Magnitude and color values for the boundaries of the intervals considered to fit the RGB of each cluster}
\label{mag_col_int}
\centering
\begin{tabular} {l c c c c}
\hline\hline
Cluster & lower mag & upper mag & lower color & upper color   \\
\hline
G78     &  $-$2.2     &  0.685      &  none       &     0.8       \\
G213    &  $-$2.2     &  0.685      &  $-$0.7       &     0.5       \\
G280    &  $-$1.8     &  0.685      &  none       &     0.6       \\
\hline
\end{tabular}
\end{table}

\subsection{Density distribution of fields stars vs. cluster stars}
\label{cluster_vs_field}

In order to  determine the actual  contribution of field stars  to the
RGB of  the cluster, we  need  to compare the density  distribution of
field stars to that of stars in  the cluster.  Figure \ref{dens} shows
the  density  distribution  of  the   clusters and  their  surrounding
field. Since  the cluster stars   have been selected  in  view of  the
robustness of their photometry, the most  central stars of the cluster
are missing. This leaves a hole in the radial distribution of stars.
The average density of stars in each annulus (associated to each cluster and to its surrounding field), as well as the surface of the cluster are given in Table \ref{sigmas}.

\subsection{Color difference histograms for clusters and surrounding fields}

\begin{figure}
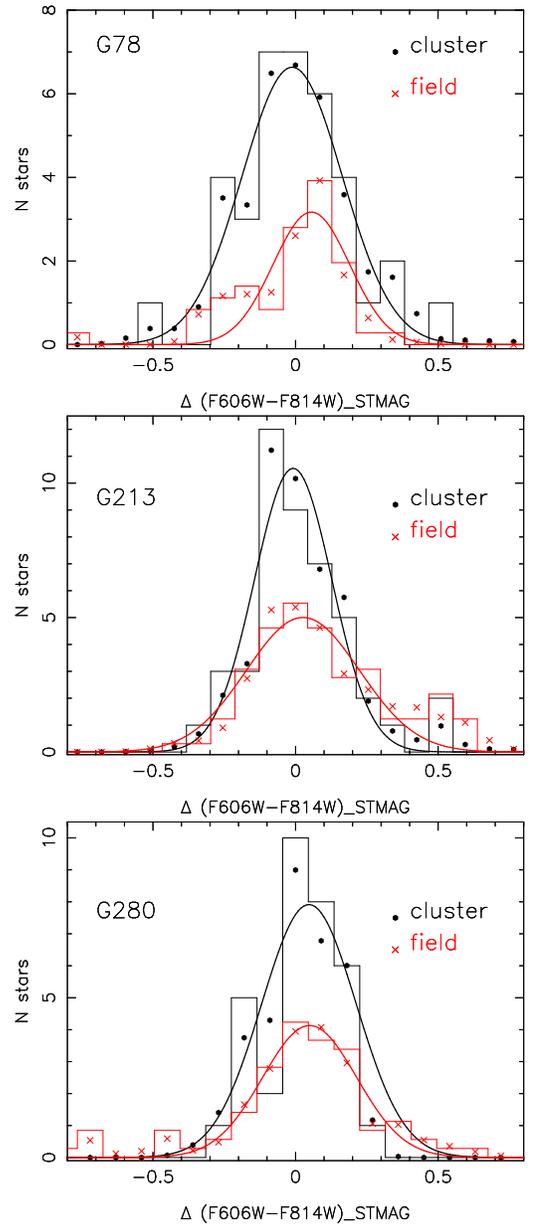

\centering
\includegraphics[scale=0.325,angle=270]{G078_hists-e022_jan08.ps}
\includegraphics[scale=0.325,angle=270]{G213_hists-e01_fev08.ps}
\includegraphics[scale=0.325,angle=270]{G280_hists-e01_fev08.ps}
\caption{Color histograms for each cluster, G78 ({\it top}), G213 ({\it middle}) and G280 ({\it bottom}), and its corresponding surrounding fields.  Solid straight lines show the simple histogram of both the cluster and the surrounding field derived by simple bining of stars per color interval. Dots (for the cluster) and crosses (for the surrounding field stars) display the histograms obtained by weighting the contribution of each star to each color bin with its associated photometric error. Solid lines indicate the Gaussian fit for each case. }
\label{hists}
\end{figure}

Figure \ref{hists} shows   the color difference histograms  around the
fiducial RGB  for each   cluster  and its  corresponding  field. These
histograms  show the  difference between  the color  of a star  with a
certain magnitude and the color value of the  fiducial RGB at the same
magnitude.  RGB stars were considered  within a magnitude interval  of
0.5 mag, with its faintest  limit set at  0.5 magnitudes above the red
clump. This value was chosen to avoid the  mixed stellar population of
the red clump, while the brighter limit of the magnitude interval (0.5
magnitudes brighter) was chosen  to avoid the upper parts of  the RGB
that   could be polluted by   AGB  stars. The  magnitude interval  for
the three    clusters goes from 0.185   to  0.685 mag.  Histograms were
derived using a color bin of 0.085 mag for G78  and G213 and 0.090 mag
for G280.  The  size  of  color bin  was  chosen  to best sample   the
distribution of stars along  this part  of the  RGB according to  each
cluster. The  distribution of the field stars  was scaled  to the same
area  as   the cluster  stars based   on  the values of   $R_{in}$ and
$R_{out}$ shown in Table \ref{sigmas}.  
Histograms were obtained  by  direct bining of stars  per
unit color  (straight line histograms   in Figure \ref{hists}).
So-called ``generalized histograms'' (filled circles in the same figure) were obtained by weighting the contribution of  each star to each color bin of the histogram with its photometric error, i.e. the value of the photometric error of each star, determined the weight this star would add to the star counts in each color bin of the histogram.  
For
all three clusters,  the  first histogram (simple histogram)  and  the
second one ({\it generalized} histogram) follow  the same trend.  Once the color distributions of the field stars had been  properly  scaled to the area considered for the determination of the color distribution of the cluster stars, the field stars were  directly   subtracted from  the  cluster   stars per  color  and
magnitude bin in order to derive a color distribution of cluster stars
along the RGB.

\section{Analysis of RGB width}
\label{RGBwidth}

\begin{figure}
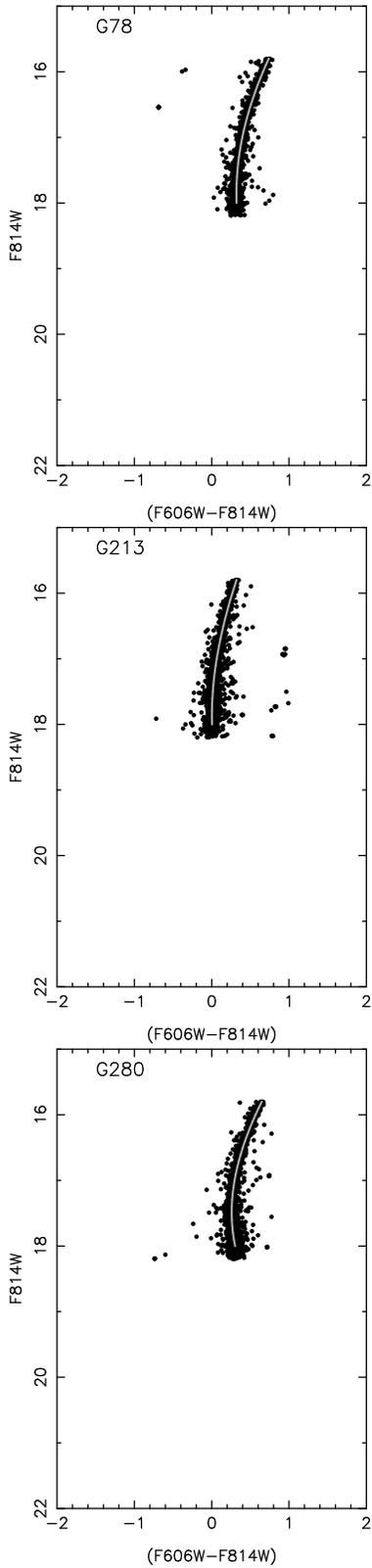

\centering
\includegraphics[scale=0.45,angle=270]{G078_ASE_statsel_jan08.ps}
\includegraphics[scale=0.45,angle=270]{G213_ASE_statsel_fev08.ps}
\includegraphics[scale=0.45,angle=270]{G280_ASE_statsel_fev08.ps}
\caption{Results of the artificial star experiments after applying the statistical selection of stars described in section \ref{stat_sel} for G78 ({\it top left}), G213 ({\it top right}) and G280 ({\it bottom}). The solid line indicates the 
input RGB while the dots indicate the output (measured)  RGB stars. Only stars falling within each cluster annulus were considered. }
\label{M31_ASE_stat}
\end{figure}

In order to evaluate  the contribution  of  photometric errors
to  the
width of each cluster's  RGB, we derived the CMD of the recovered artificial stars after applying a similar statistical selection than the one done for the observed stars and described in section \ref{stat_sel}, as well as using the magnitude and color limits shown in Table \ref{mag_col_int}. Only artificial stars falling within each cluster annulus are considered. Figure  \ref{M31_ASE_stat} shows the CMD of these artificial stars.

\begin{figure}
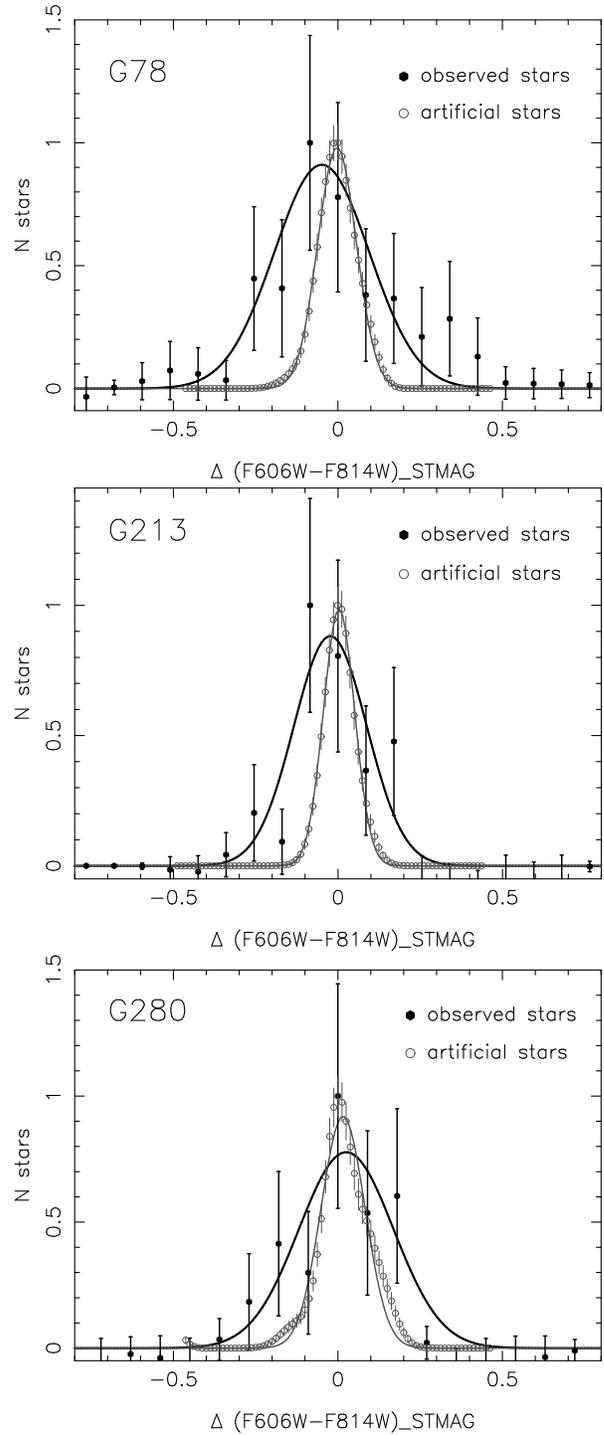

\centering
\includegraphics[scale=0.375,angle=270]{G078_final-e022_jan08.ps}
\includegraphics[scale=0.375,angle=270]{G213_final-e01_fev08.ps}
\includegraphics[scale=0.375,angle=270]{G280_final-e01_fev08.ps}
\caption{Resulting color histogram for G78 ({\it top}), G213 ({\it middle}) 
and G280 ({\it bottom}). Filled dots display the color 
distribution of observed stars. 
Empty circles display the color distribution of recovered artificial stars. Solid and 
dashed lines showed the fitted Gaussians for each case. }
\label{final}
\end{figure}

Considering the same magnitude intervals than the ones used in the previous section  to  compute the color histograms of the observed stars, we derived  the color histogram  of the recovered artificial  stars. Empty circles  in Figure \ref{final} show
the generalized normalized histogram of the recovered artificial stars
per  cluster. The dashed line  shows the Gaussian fit.  Filled circles
in the same figures  display the generalized normalized  histogram for
the  observed  stars without the contribution   of field  stars. Solid
lines  show   the Gaussian  fits.  The   resulting histograms  for the
observed stars are  not  centered on  zero  due to  the  fact that the
fiducial RGB was derived before the subtraction  of field stars, i.e.,
considering both cluster and   field stars. Once the   subtraction was
done, the locus of the remaining distribution is slightly shifted from
zero.  For G78, there  is a shift of  $-$0.03 mag; for G213, the shift
equals $-$0.03 mag; for G280, there is a shift of 0.02 mag.
Rows 6 and 7 of Table \ref{sigmas} display  the  values of $\sigma$ for the
two  Gaussian fits along  with  their uncertainties.  \ $\sigma_{obs}$
corresponds to the distribution of observed stars while $\sigma_{art}$
corresponds  to that of    the  artificial stars.  These values   were
subtracted  in quadrature in order to  analyze the  intrinsic width of
the observed stars distribution for each cluster.  The last row
of Table \ref{sigmas} show the resulting difference, $\sigma_{int} \ =
\ (\sigma^2_{obs} - \sigma^2_{art})^{1/2}$ and its uncertainty.

\begin{table*}
\caption{Parameters for the RGB width analysis and sigma values from the Gaussian fits of color distributions of observed and artificial stars.
}
\label{sigmas}
\centering
\begin{tabular} {l c c c c c c c }
\hline
\hline
Cluster ID & &  & G78  &  & G213 &  & G280  \\
\hline
$R_{in} (arcsec) $  & & & 5.1 & & 3.7 & & 4.2 \\
$R_{out} (arcsec)v$ & & & 6.4 & & 5.3 & & 5.6 \\

Cluster density (stars per arcsec$^{2}$) & & & 14 & & 20 & & 19 \\
Field density (stars per arcsec$^{2}$)   & & & 7 & & 15 & & 12 \\

Cluster surface (arcsec$^{2}$)  & & & 55 & & 45 & & 43 \\
$\sigma_{obs} \pm \Delta \sigma_{obs}$ & & & 0.1449 $\pm$ 0.0395 & & 0.1110 $\pm$ 0.0261 & & 0.1439 $\pm$ 0.0415 \\
$\sigma_{art} \pm \Delta \sigma_{art}$ & & & 0.0587 $\pm$ 0.0142 & & 0.0472 $\pm$ 0.0085 & & 0.0566 $\pm$ 0.0129 \\
$\sigma_{int} \pm \Delta \sigma_{int}$ & & & 0.1325 $\pm$ 0.0397 & & 0.1005 $\pm$ 0.0261 & & 0.1323 $\pm$ 0.0416 \\
\hline
\end{tabular}
\end{table*}

\section{Determination of the color-metallicity relation and metallicity dispersion}

\begin{figure}
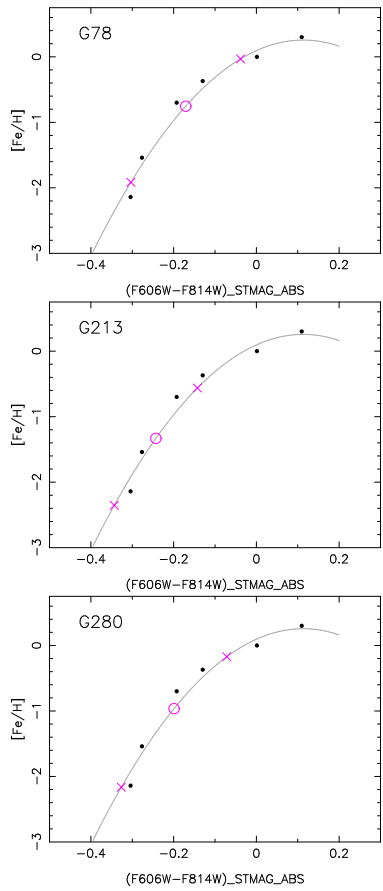

\centering
\includegraphics[scale=0.275,angle=270]{G078_colmet-e022_jan08.ps}
\includegraphics[scale=0.275,angle=270]{G213_colmet-e01_fev08.ps}
\includegraphics[scale=0.275,angle=270]{G280_colmet-e01_fev08.ps}
\caption{ Color versus metallicity plot for a magnitude value of F814W=0.435 derived from the \citet{brow05} ridgelines. Solid line shows the fit of second order polynomial given by equation (1). Empty circle shows the position of the RGB of G78 at that magnitude value considering $E(B-V)$= 0.22 ({\it top left panel}); empty circle shows the position of the RGB of G213 at that magnitude value considering a reddening of 0.1 ({\it top right panel}); empty circle shows the position of the RGB of G280 at that magnitude value considering a reddening of 0.1 ({\it bottom panel}). For all three panels, crosses indicate the lower and upper limits in color and metallicity used to determine the metallicity dispersion of each cluster. }
\label{colmet}
\end{figure}

Assuming the intrinsic color width of the RGB of these clusters is due
entirely  to a  metallicity dispersion,  we can  estimate the value of
this dispersion  by constructing  the color-metallicity relation using
the standard RGB sequences of \citet{brow05}.  Considering the central
magnitude  of the interval used  to estimate the intrinsic color width
of the clusters,  we plotted the  corresponding color values for  each
standard RGB sequence in  \citet{brow05} versus its metallicity. These
points were fitted using a second  order polynomial in order to derive
the color-metallicity  relation  for the clusters  as  shown in Figure
\ref{colmet}.  The resulting  color-metallicity relation for the three
clusters is:

$$ [Fe/H] \ = \ 0.0952 \ + \ 2.8323(V-I)_0 \ - \ 12.5153(V-I)_0^2  \eqno(1) $$

where  $(V-I)_0$ is  the RGB  color   at I=0.435  mag,   which is 0.75
magnitudes brighter than the HB.  Using this expression, we derived a
metallicity dispersion for each cluster.

First  we compute the metallicity   for the color  $(V-I)_0$, which we
shall call  $[Fe/H]_0$.  Then we add  to $(V-I)_0$ the intrinsic color
dispersion computed from the histogram  fit in section \ref{RGBwidth},
that is $\sigma_{int}$, to have  $(V-I)_0 + \sigma_{int}$.  This value
is  inserted  into equation   1 to derive   the "upper  limit"  of the
metallicity dispersion, $[Fe/H]_{up}$.   For the "lower  limit" of the
metallicity dispersion, we subtract to  $(V-I)_0$ the intrinsic  color
width,  that is $(V-I)_0 - \sigma_{int}$,  and insert  this value into
the [Fe/H] vs color relation to obtain $[Fe/H]_{low}$. These lower and
upper  limit  are indicated  with crosses in  Figure \ref{colmet}. The
metallicity dispersion will be given  by $( [Fe/H]_{up} - [Fe/H]_{low}
)  / 2 $  even though $[Fe/H]_0$  does not  lie in  the middle of this
interval.   The metallicity dispersion  of  each cluster  is given  in
Table \ref{met_disp}. We give  both  the resulting dispersion and  the
relative dispersion,   $\sigma_{[Fe/H]}  / [Fe/H]$ , where   [Fe/H] is
given in Table \ref{props_amas}.

As mentioned in Section 5, we have performed the photometry and the
artificial stars experiments with great care. The latter have been conducted in such a
way as to include all known sources of photometric error that are
readily quantified. Nonetheless,  there is a small chance
that we have underestimated the actual photometric errors, in which case,
our inferred cluster metallicity dispersions would be upper limits of the
true values.


\begin{table}
\caption{Metallicity dispersion for each cluster}
\label{met_disp}
\centering
\begin{tabular} {l c c c c c c c }
\hline
\hline
Cluster ID & &  & G78  &  & G213 &  & G280  \\
\hline
$\sigma_{[Fe/H]}$ & & & 0.9419      & & 0.8955      & & 1.0337  \\
error             & & & $\pm$0.3824 & & $\pm$0.1979 & & $\pm$0.2587 \\
 & &  &   &  &  &  &   \\
$\sigma_{[Fe/H]} / [Fe/H]$ & & & 1.0238      & & 0.8292      & & 1.4767  \\
error                      & & & $\pm$0.4157 & & $\pm$0.1832 & & $\pm$0.3695 \\
\hline
\end{tabular}
\end{table}

As a    comparison,   the  intrinsic metallicity   dispersion    of G1
($\sigma=25.1$  km s$-1$,  [Fe/H]=$-$   1.08) is  $\pm  0.5$  dex  for
$E(B-V)=0.10$, and $\pm$0.39 dex for $E(B-V)=0.05$ \citep{meylan01}, which is equivalent to
that of the three clusters within  the error bars. We note that the
G280's  metallicity dispersion  that we  derive  in this work is  much
higher   than the value  found     by \citep{steph01} in their   study
involving NICMOS observations of  metal-rich globular clusters in M31.
According   to those authors, $\sigma_{[Fe/H]}  \sim   0.2$, using the
dispersion   in  the  measured   $(J-H)$  and  $(J-K)$  colors.   This
difference is probably due to the fact that for  any spread in [Fe/H],
the corresponding spread  in color is twice  as small in the  infrared
$(J-K)$ than in the visible $(V-I)$. Also the fact that they work with
a smaller sample  ($\sim$200 stars {\it  versus} $\sim$2500 stars  per
cluster in this work)  would statistically affect their measurement of
any spread in metallicity.

\section{Discussion and Conclusions}

In light  of  the existing  hypotheses made  for  the case   of our
Galaxy, we now consider some possible explanations for the metallicity
dispersion seen in the four most massive globular clusters in M31.

In  a scenario of  chemical  self-enrichment, the cluster  metallicity
dispersion comes from stars originating  from  the gas trapped in  the
globular  cluster after its formation.   Therefore, one expects a more
massive cluster to  retain more gas and  thus to exhibit a more
prominent contribution from the second generation of stars.  
In our case, the metallicity dispersions of the three clusters are similar within the error bars. On the other hand, if their velocity dispersions are directly linked to their masses, G78 and G280 have similar masses, while G213 is less massive. With this information, we are not able to establish any link between the metallicity dispersion and the mass of the clusters. This  lack of a correlation between  the cluster metallicity   dispersion and velocity  dispersion could indicate that the role of the cluster mass in widening its RGB is not straight-forward. However it is hard to draw any conclusion from such a small sample of clusters.

A second attractive explanation for  the observed width  of the RGB of
these clusters would be that they are  actually the remains of tidally
stripped dwarf galaxies.  Recent  results on dwarf-globular transition
objects  (DGTOs -Hasegan et al.   2005)  and ultra-compact dwarf (UCD)
galaxies in Virgo  \citep{jones06} and Fornax  \citep{drink03} support
the hypothesis  that these objects are  the remnant nuclei of threshed
dE,N galaxies.  Zinnecker   et al.   (1988)  and Freeman  (1993)  have
already suggested the possibility of $ \omega$ Cen being the remaining
core of  a larger entity  such as a  former nucleated dwarf elliptical
galaxy.   \citet{bekki03}  suggest that the  thick  disk of our Galaxy
might  have been formed by the  accretion of dwarf spheroidals such as
$\omega$Cen.  The results of  \citet{meylan01}   on G1 indicate   this
cluster could be a  kind of transition  step between globular clusters
and  dwarf elliptical galaxies,  being  the remaining core  of a dwarf
galaxy whose envelope would   have   been severely pruned   by   tidal
shocking due to     the bulge and    disk of  its host galaxy,    M31.
Considering that  M31 also possesses a   thick disk \citep{ata01}
and  seems to have suffered more tidal stripping events than the MW
\citep{ibata01,brow06},  one   would  expect  the  presence  of  other
remaining  cores as ``relics'' of the  history of interactions of this
galaxy resembling  the clusters studied  in this  work. In fact, both
G213  and G280 seem to fall  (at least  in projection)  on  one of the
trajectories   suggested by  \citet{ibata04}   for the   passage of  a
satellite galaxy that might have been tidally disrupted by M31.

In  order to explore   this idea, we   searched for any tidal features
(tails or asymmetric envelopes) around the clusters. We looked for any
density enhancements in  the outer parts  of each cluster to the edges
of the FoV. For all three clusters there is no clear density excess in
any particular direction that could indicate the presence of any tidal
feature. However this  visible absence of  density enhancement is not
conclusive,  since at the  distance of M31, it   would be difficult to
detect any  density enhancements from the  star-counts of the brighter
stars  of the cluster -contrary to  density  enhancement detections in
GCs in our Galaxy through star-counting methods including stars on the
main sequence (i.e. Grillmair et al. 1995; Lee et al. 2003).

A third possibility would be a primordial metallicity inhomogeneity in
the proto-cluster cloud induced by the influence of field stars during
the formation  of  the globular cluster.    For a  long  time the main
question was whether these inhomogeneities were inherited at the birth
of  the  stars    that  we are  currently    observing  (the so-called
self-enrichment hypothesis  that implies pollution of the intracluster
gas  by more massive   and faster  evolving  stars) or  if  they  were
generated   in  the course  of the    evolution of these  objects (the
so-called   evolution  hypothesis  that  requires non-canonical mixing
inside  the   star  itself).   On    the  one  hand,   the   classical
self-enrichment scenario (Truran et al.  1991; Prantzos
\& Charbonnel 2006 -and references therein) assumes that stars within
the  GCs created  the  totality of the  metals  (light and heavy), and
involves  three  stellar   generations.    On  the  other  hand,   the
pre-enrichment scenario   \citep{kraft94} assumes that GCs   were born
with  their current heavy  metal content and  invokes only two stellar
generations.    Both   hypotheses  are    globally supported   by  our
observations.  In   order to  really    be able  to  have   conclusive
arguments, we would need detailed  chemical information (on $\alpha$-,
r-  and  s-process elements  for example),   while the color-magnitude
diagnostics only give information on  the cluster global metallicities
(Z).

In all of the above scenarios, the  realm of parameter space for these
clusters  is large and   we are  only  beginning to  explore  what the
significant metallicity   dispersion   among their  stars   might   be
correlated with. In conclusion, the results presented in this work add
to the growing body of evidence  that massive globular clusters have a
more complex  history of star  formation and chemical  enrichment than
would   be  expected in a  traditional   view of  globular clusters as
simple,  single-age  and single-metallicity  stellar systems.  Further
studies of    these  and similar   objects  will   likely  expand  our
understanding of galaxy assembly.

\begin{acknowledgements}

IFC acknowledges the financial  support of FAPESP  grant no.03/01625-2
and the  Sixth Program of  the EU  for a  Marie Curie Fellowship. 
AS was partially supported by STScI grant HST-GO-09719.01-A.
SGD acknowledges a partial support from the STScI grant HST-GO-09719.03-A,
and  the  Ajax  Foundation. IFC   also  thanks D.R. Gon\c calves and
H. Flores for computational support.

\end{acknowledgements}


\begin{thebibliography}{}

\bibitem[Barmby et al(2000)]{barm00} Barmby, P.,Huchra, J.P., Brodie, J.P., Forbes, D.A., Schroder, L.L. \& Grillmair, C.J. 2000, \aj, 119, 727 

\bibitem[Bekki \& Freeman(2003)]{bekki03} Bekki, K. \& Freeman, K.C. 2003, \mnras, 346, L11

\bibitem[Bekki et al(2007)]{bekki07} Bekki, K.,Campbell, S.W., Lattanzio, J.C. \& Norris, J.E. 2007, \mnras, 377, 335

\bibitem[Bekki(2006)]{bekki06} Bekki, K. 2006, \mnras, 367, L24

\bibitem[Brown et al(2003)]{brow03}  Brown, T.M., Ferguson, H.C., Smith, E., Kimble, R.A., Sweigart, A.V., Renzini, A., Rich, R.M. \& VandenBerg, D.A. 2003, \apj, 592, L17

\bibitem[Brown et al(2005)]{brow05}  Brown, T.M., Ferguson, H.C., Smith, E., Guhathakurta, P.,  Kimble, R.A., Sweigart, A.V., Renzini, A.,  Rich, R.M.  \&  VandenBerg, D. A. 2005, \apj, , 1693

\bibitem[Brown et al(2006)]{brow06}  Brown, T.M.,  Smith, E., Ferguson, H.C.,  Rich, R.M., Guhathakurta, P., Renzini, A., Sweigart, A.V. \& Kimble, R.A. 2006, \apj, 652, 323	

\bibitem[Dopita \& Smith(1986)]{dopita86} Dopita, M.~A., \& 
Smith, G.~H.\ 1986, \apj, 304, 283 

\bibitem[Djorgovski \& Meylan(1994)]{djor94} Djorgovski, S.G. \& Meylan, G. 1994, \aj, 108, 1292

\bibitem[Djorgovski et al(1997)]{djor97} Djorgovski, S.G., Gal, R.R., McCarthy, J.K, Cohen,J.G, de Carvalho, R.R., Meylan, G., Bendinelli, O. \& Parmeggiani, G. 1997, \apj, 474, L19

\bibitem[Drinkwater et al(2003)]{drink03} Drinkwater, M.J., Gregg, M.D., Hilker, M.J., Bekki, K., Couch, W.J., Ferguson, H.C., Jones, J.B. \& Philipps, S. 2003, Nature, 423, 519 

\bibitem[Ferguson et al(2006)]{fer06} Ferguson, A., Irwin, M., Chapman, S., Ibata, R., Lewis, G. \& Tanvir, N. astro-ph/0601121

\bibitem[Freedman \& Madore(1990)]{freed90} Freedman, W.L. \& Madore, B.F 1990, ApJ, 365, 186 

\bibitem[Freeman(1993)]{free93} Freeman, K.C. 1993, \pasp, 48, 608

\bibitem[Frogel, Persson \& Cohen(1980)]{frog80} Frogel, J.A., Persson, S.E. \& Cohen, J.G.
1980, \apj, 240, 785

\bibitem[Gratton et al.(2004)]{gratton2004} Gratton, R., Sneden, C., \& Carretta, E.\ 2004, \araa, 42, 385 

\bibitem[Grillmair et al(1995)]{grill95} Grillmair, C.J., Freeman, K.C., Irwin, M. \& Quinn, P.J. 1995, \aj, 106, 2553  

\bibitem[Hasegan et al(2005)]{has05} Hasegan, M., Jord\'an, A., C\^o t\'e, P., Djorgovski, S.G., McLaughlin, D.E., Blakeslee, J.P., Mei, S., West, M.J., Peng, E.W., Ferrarese, L., Milosavljevic M., Tonry, J.L. \& Merritt, D. 2005, \apj, 627, 203

\bibitem[Hilker \& Richtler(2000)]{hilker2000} Hilker, M., \& Richtler, T.\ 2000, \aap, 362, 895 

\bibitem[Hodge(1981)]{hodge81} Hodge, P.W. 1981, Atlas of the Andromeda Galaxy, University of Washington Press 

\bibitem[Huchra, Brodie \& Kent(1991)]{huch91} Huchra, J.P., Brodie, J.P. \& Kent, S.M. 1991, \apj, 370, 495

\bibitem[Ibata et al(2001)]{ibata01} Ibata, R., Irwin, M., Lewis, G., Ferguson, A.M.N. \& Tanvir, N. 2001, Nature, 412, 49

\bibitem[Ibata et al(2004)]{ibata04} Ibata, R., Chapman, S., Ferguson, A.M.N., Irwin, M., Lewis, G. \& McConnachie, A. 2004, \mnras, 351, 117

\bibitem[Ikuta \& Arimoto(2000)]{iku00} Ikuta, C. \& Arimoto, N. 2000, \aap, 358, 535

\bibitem[Jablonka, Alloin \& Bica(1992)]{pj92} Jablonka, P., Alloin, D. \& Bica, E. 1992, \aap, 260, 97

\bibitem[Jones et al(2006)]{jones06} Jones, J.B., Drinkwater, M.J., Jurek, R., Philipps, S., Gregg, M.D., Bekki, K. \& Couch, W.J. 2006, \aj, 131, 312

\bibitem[Koekemoer et al(2002)]{koek02} Koekemoer, A.M., Fruchter, A.S., Hook, R. \& Hack, W. 2002, HST Calibration Workshop, eds. S. Arribas, A. Koekemoer \& B. Whitmore, 337

\bibitem[Koornef et al(1986)]{koor86} Koornef, J., Bohlin, R., Buser, R., Horne, K. \& Turnsheck, D. 1986, in Proceedings of the IAU 19th General Assembly, ed. S-P Swinggs

\bibitem[Kraft(1994)]{kraft94} Kraft, R.P. 1994, PASP, 106, 553

\bibitem[Lee et al.(1999)]{lee99} Lee, Y.-W., Joo, J.-M., Sohn, Y.-J., Rey, S.-C., Lee, H.C., \& Walker, A.~R.\ 1999, \nat, 402, 55 

\bibitem[Lee et al(2003)]{lee03} Lee, K.H., Lee, H.M., Fahlman, G.G. \& Lee M.G. 2003, \aj, 126, 815 

\bibitem[Meylan et al(1995)]{meylan95} Meylan, G., Mayor, M., Duquennoy, A. \& Dubath, P. 1995, \aap, 303, 761

\bibitem[Meylan et al(2001)]{meylan01} Meylan, G., Sarajedini, A., Jablonka, P., Djorgovski, S.G., Bridges, T. \& Rich, R.M. 2001, \aj, 122, 830

\bibitem[Norris \& Da Costa(1995)]{norris95} Norris, J.E. \& Da Costa, G.S. 1995, \apj, 447, 680

\bibitem[Prantzos \& Charbonnel(2006)]{pran06} Prantzos N. \& Charbonnel C. 2006, \aa, 458, 135

\bibitem[Sarajedini \& Van Duyne(2001)]{ata01} Sarajedini, A. \& Van Duyne, J. 2001, AJ, 122, 2444

\bibitem[Sarajedini et al(2006)]{ata06} Sarajedini, A., Barker, M.K., Geisler, D., Harding, P. \&  Schommer, R. 2006, \aj, 132, 1361

\bibitem[Sirianni et al(2005)]{sir05} Sirianni, M., Jee, M.J., Benitez, N., Blakeslee, J.P., Martel, A.R. et al
2005, PASP, 117, 1049

\bibitem[Smith \& McClure(1987)]{smith87} Smith, G.H. \& McClure, R.D. 1987, \apj, 316, 206 
 
\bibitem[Stephens et al(2001)]{steph01} Stephens, A.W., Frogel, J.A., Freedman, W., Gallart, C., Jablonka, P., Ortolani, S., Renzini, A., Rich, R.M. \& Davies, R. 2001, \aj, 121, 2597

\bibitem[Stetson(1994)]{stet94} Stetson, P.B. 1994, PASP, 106, 250

\bibitem[Tsujimoto \& Shigeyama(2003)]{tsuji03} Tsujimoto, T., \& Shigeyama, T.\ 2003, \apj, 590, 803 

\bibitem[Trutan et al(1991)]{tru91} Truran, J.W., Brown, J. \& Burkert, A. 1991, ASPC, 13, 78

\bibitem[Zinnecker et al(1988)]{zin88} Zinnecker, H., Keable, C.J., Dunlop, J.S., Cannon, R.D. \& Griffiths, W.K. 1988, IAUS, 126, 603 


\end{thebibliography}
\end{document}